\documentstyle[12pt,epsfig]{article}

\newcommand{\ba}{\begin{eqnarray}}
\newcommand{\ea}{\end{eqnarray}}
\begin{document}

\title{Geometrical analysis of the algebraic cluster model}

\author{R. Bijker\\
Instituto de Ciencias Nucleares,\\
Universidad Nacional Aut\'onoma de M\'exico,\\
A.P. 70-543, 04510 M\'{e}xico D.F., M\'{e}xico}
\date{}
\maketitle

\begin{abstract}

Three-body clusters are studied in the algebraic cluster model. 
Particular attention is paid to the case of three identical particles. 
It is shown in a geometrical analysis that the harmonic oscillator, the 
deformed oscillator and the oblate symmetric top are contained as special 
solutions. An application of the oblate top limit to the nucleus $^{12}$C 
suggests that the low-lying spectrum can be described as a configuration 
of three identical $\alpha$ particles at the vertices of an equilateral 
triangle.   

\

Se investigan c\'umulos de tres part{\'{\i}}culas id\'enticas en un modelo 
algebraico de c\'umulos. Se muestra expl{\'{\i}}citamente que el oscilador 
arm\'onico, el oscilador deformado y el trompo oblato corresponden a 
soluciones especiales del modelo. La aplicaci\'on del trompo oblato al 
n\'ucleo $^{12}$C sugiere que los niveles a bajas energ{\'{\i}}as pueden 
describirse como una configuraci\'on de tres part{\'{\i}}culas $\alpha$ 
localizadas a los v\'ertices de un tri\'angulo equil\'atero.

\

\noindent
PACS numbers: 21.60.Fw, 21.60.Gx, 27.20.+n

\end{abstract}

\section{Introduction}

Algebraic models have found useful applications both in many-body and in 
few-body systems. In general terms, in algebraic models energy 
eigenvalues and eigenvectors are obtained by diagonalizing a 
finite-dimensional matrix, rather than by solving a set 
of coupled differential equations in coordinate space. As an example we 
mention the interacting boson model (IBM), which has been very 
successful in the description of collective states in nuclei \cite{IBM}. 
Its dynamical symmetries correspond to the quadrupole vibrator, the axially 
symmetric rotor and the $\gamma$-unstable rotor in a geometrical description. 
In addition to these special solutions, the IBM can describe intermediate 
cases between any of them equally well. 
The first application of algebraic models to few-body systems was the 
vibron model \cite{vibron}, which was introduced to describe vibrational and 
rotational excitations in diatomic molecules. The dynamical symmetries of the 
vibron model correspond to the (an)harmonic oscillator and the Morse 
oscillator. 

The principal idea is to introduce a $U(k+1)$ spectrum generating algebra 
for a problem of $k$ degrees of freedom. The $k=5$ quadrupole degrees of 
freedom in collective nuclei thus leads to the $U(6)$ interacting boson model, 
and the $k=3$ dipole degrees of freedom of the relative coordinate in the 
two-body problem to the $U(4)$ vibron model. For three particles we recover 
the $U(7)$ model which was developed originally to describe the relative 
motion of the three constituent quarks in baryons \cite{BIL}. 

The aim of this contribution is to study three-body clusters in nuclear 
physics. First we discuss some special solutions of the Schr\"odinger 
equation in coordinate space for the case of three identical particles. 
Next we introduce a $U(7)$ interacting boson model for the relative 
motion of three clusters: the Algebraic Cluster Model (ACM). Its algebraic 
properties are interpreted geometrically with mean-field methods, which 
use coherent states, classical limits and Bohr-Sommerfeld 
requantization techniques. It is shown the ACM Hamiltonian contains the 
spherical oscillator, the deformed oscillator and the oblate top as special 
cases. Finally, we study an application to three-alpha configurations in 
nuclei, in particular to the energy spectrum and form factors of $^{12}$C. 

\section{Integro-differential methods}

\begin{figure}
\centering
\setlength{\unitlength}{1.0pt}
\begin{picture}(200,135)(0,0)
\thicklines
\put ( 25, 10) {\circle*{10}}
\put (125, 10) {\circle*{10}}
\put ( 75,110) {\circle*{10}}
\put ( 25, 10) {\line ( 1,0){100}}
\put ( 25, 10) {\line ( 1,2){ 50}}
\put (125, 10) {\line (-1,2){ 50}}
\put ( 25, 10) {\line ( 3, 2){ 50}}
\put (125, 10) {\line (-3, 2){ 50}}
\put ( 75,110) {\line ( 0,-1){ 67}}
\put ( 75,120) {1}
\put ( 10,  0) {2}
\put (135,  0) {3}
\end{picture}
\caption[]{Geometry of a three-body system.}
\label{geometry}
\end{figure}
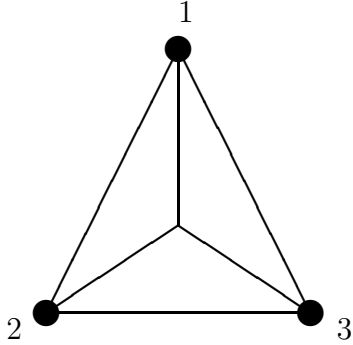

The quantum treatment of an identical three-body cluster can be done in
several ways. In \cite{greiner} the starting point is a classical Hamiltonian 
which is subsequently quantized using the Pauli-Podolsky method. Here instead 
we use a quantum mechanical treatment from the outset \cite{ACM}. First 
we introduce Jacobi coordinates 
\ba
\vec{\rho} &=& \left( \vec{r}_{1} - \vec{r}_{2} \right) /\sqrt{2} ~,  
\nonumber \\
\vec{\lambda} &=& \left( \vec{r}_{1} + \vec{r}_{2} - 2\vec{r}_{3} \right)/
\sqrt{6} ~,  
\label{Jacobi}
\ea
to describe the geometric configuration of Fig.~\ref{geometry}. 
Next we write down a Hamiltonian in terms of these coordinates and their 
canonically conjugate momenta $\vec{p}_{\rho}$ and $\vec{p}_{\lambda}$, 
and solve the Schr\"odinger equation 
\ba
\left[ \frac{1}{2m}(\vec{p}_{\rho }^{\,2}+\vec{p}_{\lambda }^{\,2}) 
+ V(\vec{\rho},\vec{\lambda}) \right] \psi (\vec{\rho},\vec{\lambda})
&=& E \, \psi (\vec{\rho},\vec{\lambda}) ~. 
\ea
In order to obtain the energy eigenvalues and eigenvectors, it is convenient 
to make a change of variables from $\vec{\rho}$, $\vec{\lambda}$ to the 
hyperradius $r$, the hyperangle $\xi$, the relative angle $2\theta$ 
\ba
\rho \;=\; r\sin \xi ~, \hspace{1cm} \lambda \;=\; r\cos \xi ~, 
\hspace{1cm} \cos 2\theta \;=\; \hat{\rho}\cdot \hat{\lambda} ~, 
\ea
and the three Euler angles $\Omega$ of the body-fixed frame. In general, a 
rotationally invariant potential only depends on the intrinsic variables 
$r$, $\xi$ and $\theta$. 

\subsection{Hyperspherical potentials}
\label{hypot}

For potentials that only depend on the hyperradius $r$, the Schr\"odinger
equation can be solved by separation of variables into an angular and a
radial equation. For the six-dimensional harmonic oscillator 
\ba
V(r) &=& \frac{1}{2} C \, r^{2} ~,
\ea
the energy spectrum can be obtained exactly as 
\ba
E(n) &=& \epsilon \, (n+3) \;=\; \epsilon \, (2n_r+\sigma+3) ~, 
\label{ener1}  
\ea
with $n=0,1,\ldots$~, and $\epsilon =\sqrt{C/m}$. The allowed values of 
$\sigma$ are $\sigma=n,n-2,\ldots,1$ or $0$ for $n$ odd or even, 
respectively. The radial quantum number $n_r$ can have $n_r=0,1,\ldots$ ~.  

For the six-dimensional displaced (or deformed) oscillator 
\ba
V(r)&=&\frac{1}{2}C(r-r_{0})^{2}~,
\ea
the energy eigenvalues can be obtained in closed form 
in the limit of small oscillations around the equilibrium value 
$r_{0}$ \cite{ACM} 
\ba
E(v,\sigma) &\cong& \epsilon \, (v+\frac{1}{2})
+\frac{1}{2mr_{0}^{2}} \left[ \sigma (\sigma +4)+\frac{15}{4}\right] ~,
\label{ener2}
\ea
with $\epsilon=\sqrt{C/m}$. The first term gives rise to a harmonic 
vibrational spectrum with $v=0,1,\ldots$~, whereas the second term gives the 
rotational spectrum with $\sigma =0,1,\ldots$~. 

\subsection{Spherical potentials}

In general, the potential is not invariant under six-dimensional rotations
as in the previous two examples, but only under rotations in three
dimensions. An interesting situation occurs when the potential has sharp
minima both in $r$, $\xi$ and $\theta$ 
\ba
V(r,\xi,\theta) &=& \frac{1}{2} C (r-r_{0})^{2} 
+ \frac{1}{2} A (\xi-\xi_{0})^{2} 
+ \frac{1}{2} B (\theta-\theta_{0})^{2} ~.
\ea
In the limit of small oscillations around $r_{0}$, $\xi_{0}=\pi/4$ and 
$\theta_{0}=\pi/4$, rotations and vibrations decouple, and 
the set of resulting differential equations can be solved in closed form. 
For the case of three identical clusters, the potential has to be invariant 
under their permutation, i.e. the coefficients $A$ and $B$ are equal. 
The energy spectrum is then given by 
\ba
E(v_{1},v_{2},I,K) &\cong& \epsilon_{1} \, (v_{1}+\frac{1}{2}) 
+ \epsilon_{2} \, (v_{2}+1) 
+ \frac{1}{mr_{0}^{2}} \left[ I(I+1) - \frac{1}{2}K^{2} 
-\frac{9}{8} \right] ~,  \label{ener3} 
\ea
with $\epsilon_{1}=\sqrt{C/m}$, $\epsilon_{2}=\sqrt{A/mr_{0}^{2}}$. 
Here $K$ is the projection of the angular momentum $I$ on the symmetry axis 
(perpendicular to the $\vec{\rho}$-$\vec{\lambda}$ plane). The first two 
terms in Eq.~(\ref{ener3}) describe the vibrational excitations of an 
oblate symmetric top, and the last term the rotational excitations of each 
vibrational band. 

\section{The Algebraic Cluster Model}

In this section, we introduce the Algebraic Cluster Model (ACM) as an 
algebraic treatment of three-cluster systems in which the eigenvalues are 
obtained by matrix diagonalization instead of by solving a set of 
differential equations. The ACM is an interacting boson model to describe 
the relative motion of the three clusters. The method consists in 
quantizing the Jacobi coordinates and momenta of Eq.~(\ref{Jacobi}) with 
vector boson creation and annihilation operators and adding an additional 
scalar boson \cite{BIL} 
\ba
b_{\rho,m}^{\dagger} \;,\; b_{\lambda,m}^{\dagger} \;,\; s^{\dagger}
\hspace{1cm} (m=0,\pm 1) 
\ea
The set of 49 bilinear products of creation and annihilation operators 
spans the Lie algebra of $U(7)$. All operators, such as the Hamiltonian and 
electromagnetic transition operators, are expanded into elements of this 
algebra. The Hamiltonian generally includes up to two-body interaction terms 
that, by construction, commute with the number operator
\ba
\hat{N} &=& s^{\dagger}s + \sum_{m} \left( b_{\rho,m}^{\dagger} b_{\rho,m} 
+ b_{\lambda,m}^{\dagger} b_{\lambda,m} \right) ~. 
\label{number}
\ea
The most general one- and two-body Hamiltonian to describe the relative 
motion of a system of three identical clusters is given by \cite{ACM} 
\ba
H &=&\epsilon _{0}\,s^{\dagger }\tilde{s}-\epsilon _{1}\,(b_{\rho }^{\dagger
}\cdot \tilde{b}_{\rho }+b_{\lambda }^{\dagger }\cdot \tilde{b}_{\lambda
})+u_{0}\,(s^{\dagger }s^{\dagger }\tilde{s}\tilde{s})-u_{1}\,s^{\dagger
}(b_{\rho }^{\dagger }\cdot \tilde{b}_{\rho }+b_{\lambda }^{\dagger }\cdot 
\tilde{b}_{\lambda })\tilde{s}  \nonumber \\
&&+v_{0}\,\left[ (b_{\rho }^{\dagger }\cdot b_{\rho }^{\dagger }+b_{\lambda
}^{\dagger }\cdot b_{\lambda }^{\dagger })\tilde{s}\tilde{s}+s^{\dagger
}s^{\dagger }(\tilde{b}_{\rho }\cdot \tilde{b}_{\rho }+\tilde{b}_{\lambda
}\cdot \tilde{b}_{\lambda })\right]  \nonumber \\
&&+\sum_{l=0,2}w_{l}\,(b_{\rho }^{\dagger }\times b_{\rho }^{\dagger
}+b_{\lambda }^{\dagger }\times b_{\lambda }^{\dagger })^{(l)}\cdot (\tilde{b%
}_{\rho }\times \tilde{b}_{\rho }+\tilde{b}_{\lambda }\times \tilde{b}%
_{\lambda })^{(l)}  \nonumber \\
&&+\sum_{l=0,2}c_{l}\,\left[ (b_{\rho }^{\dagger }\times b_{\rho }^{\dagger
}-b_{\lambda }^{\dagger }\times b_{\lambda }^{\dagger })^{(l)}\cdot (\tilde{b%
}_{\rho }\times \tilde{b}_{\rho }-\tilde{b}_{\lambda }\times \tilde{b}%
_{\lambda })^{(l)} \right.
\nonumber\\
&& \hspace{1cm} \left. +4\,(b_{\rho }^{\dagger }\times b_{\lambda }^{\dagger
})^{(l)}\cdot (\tilde{b}_{\lambda }\times \tilde{b}_{\rho })^{(l)}\right] 
+c_{1}\,(b_{\rho }^{\dagger }\times b_{\lambda }^{\dagger })^{(1)}\cdot (%
\tilde{b}_{\lambda }\times \tilde{b}_{\rho })^{(1)}~,  \label{hs3}
\ea
with $\tilde{b}_{\rho ,m}=(-1)^{1-m}b_{\rho ,-m}$, $\tilde{b}_{\lambda
,m}=(-1)^{1-m}b_{\lambda ,-m}$ and $\tilde{s}=s$. In addition to the total 
number of bosons $N$, the angular momentum $L$ and parity $P$, the wave 
functions are characterized by their transformation property under the 
permutation group $S_3$: $t=S$ for the symmetric, $t=A$ for the antisymmetric 
or $t=M$ for the mixed symmetry representation. Since we do not consider 
internal excitations of the clusters, the three-body wave function arises  
solely from the relative motion. Hence the permutation symmetry of the 
$U(7)$ wave function has to be symmetric $t=S$. 

\section{Geometrical analysis}

The geometric properties of the algebraic Hamiltonian of Eq.~(\ref{hs3}) 
can be studied with time-dependent mean-field approximations. The mean-field 
equations can be derived by minimizing the action \cite{onno,Levit}
\ba
S \;=\; \int_0^T dt \left< N;\vec{\alpha}_{\rho},\vec{\alpha}_{\lambda} 
\right| i \frac{\partial}{\partial t} - H 
\left| N;\vec{\alpha}_{\rho},\vec{\alpha}_{\lambda} \right> ~. 
\ea 
Here we have introduced an intrinsic or coherent state as a variational wave 
function for the three-body system   
\ba
|N;\vec{\alpha}_{\rho },\vec{\alpha}_{\lambda }\rangle \;=\; 
\frac{1}{\sqrt{N!}} (b_{c}^{\dagger})^{N}\,|0\rangle ~. 
\ea
The condensate boson $b_c^{\dagger}$ can be parametrized in terms 
of six complex variables as \cite{ACM}
\ba
b_{c}^{\dagger }=\sqrt{1-\vec{\alpha}_{\rho }\cdot \vec{\alpha}_{\rho
}^{\,\ast }-\vec{\alpha}_{\lambda }\cdot \vec{\alpha}_{\lambda }^{\,\ast }}%
\;s^{\dagger }+\vec{\alpha}_{\rho }\cdot \vec{b}_{\rho }^{\,\dagger }+\vec{%
\alpha}_{\lambda }\cdot \vec{b}_{\lambda }^{\,\dagger }~.
\label{bc}
\ea
The variational principle $\delta S=0$ gives Hamilton's 
equations of motion 
\ba
\dot{\pi}_{j} \;=\; -\frac{\partial H_{\rm cl}}{\partial \alpha_{j}} ~,
\hspace{1cm}
\dot{\alpha}_{j} \;=\;  \frac{\partial H_{\rm cl}}{\partial \pi_{j}} ~,
\label{eqmot}
\ea
where $\alpha_j$ and $\pi_j=i\alpha^{\ast}_j$ represent canonical coordinates 
and momenta. $H_{\rm cl}$ denotes the classical limit of the Hamiltonian. 
It is given by the coherent state expectation 
value of the normal ordered operator divided by $N$ 
\ba
H_{\rm cl} &=& \frac{1}{N} 
\langle N;\vec{\alpha}_{\rho},\vec{\alpha}_{\lambda } \mid \,: H :\, 
\mid N;\vec{\alpha}_{\rho},\vec{\alpha}_{\lambda} \rangle ~.
\ea
Bound states now correspond to periodic classical trajectories $\alpha_j(t)$, 
$\pi_j(t)$ with period $T$ that satisfy a Bohr-Sommerfeld type quantization 
rule \cite{onno}
\ba
N \int_0^T \pi_j \dot{\alpha}_j dt 
\;=\; N \oint \pi_j d\alpha_j \;=\; 2\pi n_j ~.
\ea
The energy associated with a periodic classical orbital is independent of 
time and is given by $E/N=H_{\rm cl}(\alpha_j,\pi_j)$.

For the geometrical analysis of the ACM Hamiltonian it is convenient to use 
spherical rather than cartesian coordinates and momenta. We transform the two 
vectors $\vec{\alpha}_{\rho}$ and $\vec{\alpha}_{\lambda}$ to intrinsic 
coordinates $(q_{\rho}, \theta_{\rho}, \phi_{\rho})$ and 
$(q_{\lambda}, \theta_{\lambda}, \phi_{\lambda})$ and
their conjugate momenta \cite{onno} 
\ba
\alpha _{k,\mu }=\frac{1}{\sqrt{2}}\sum_{\nu }{\cal D}_{\mu \nu }^{(1)}(\phi
_{k},\theta _{k},0)\,\beta _{k,\nu }~,
\ea
with 
\ba
\left( 
\begin{array}{c}
\beta _{k,1} \\ 
\beta _{k,0} \\ 
\beta _{k,-1}
\end{array}
\right) =\left( 
\begin{array}{c}
\,[-p_{\phi _{k}}/\sin \theta _{k}-ip_{\theta _{k}}]/q_{k}\sqrt{2} \\ 
q_{k}+ip_{k} \\ 
\,[-p_{\phi _{k}}/\sin \theta _{k}+ip_{\theta _{k}}]/q_{k}\sqrt{2}
\end{array}
\right) ~,
\ea
with $k=\rho$, $\lambda $. 

\section{Special solutions}

The Algebraic Cluster Model has a rich algebraic structure, which includes 
both continuous and discrete symmetries. It is of general interest to study 
limiting situations, in which the energy spectra can be obtained in closed 
form. These special cases correspond to dynamical symmetries of the $U(7)$ 
Hamiltonian. We first consider two dynamical symmetries of the $S_3$ 
invariant Hamiltonian, which are shown to correspond to the six-dimensional 
spherical and deformed oscillators discussed in the Schr\"odinger picture 
in section~\ref{hypot}. 

\subsection{Dynamical symmetries: the U(6) limit}

For $v_{0}=0$ in Eq.~(\ref{hs3}), we recover the six-dimensional oscillator, 
since there is no coupling between different harmonic oscillator 
shells. The oscillator is harmonic if all terms, except $\epsilon_0$ and 
$\epsilon_1$, are set to zero; otherwise it is anharmonic. This dynamical 
symmetry corresponds to the reduction 
\ba
U(7) \supset U(6) \supset SO(6) \supset \cdots 
\ea
We consider the one-body Hamiltonian 
\ba
H_1 \;=\; \epsilon_{1} \sum_{m} ( b_{\rho,m}^{\dagger} b_{\rho,m} 
+ b_{\lambda,m}^{\dagger} b_{\lambda,m} ) ~, 
\label{ham1}
\ea
whose eigenvalues are those of a six-dimensional spherical oscillator 
\ba
E_1 \;=\; \epsilon_{1} \, n ~. 
\label{e1}
\ea
The label $n$ represents the total number of oscillator quanta 
$n=n_{\rho }+n_{\lambda }=0,1,\ldots ,N$. 
This special case is called the $U(6)$ limit.  

In Fig.~\ref{hosc} we show the structure of the spectrum of the spherical
harmonic oscillator with $U(6)$ symmetry. For three identical clusters, the
physical wave functions transform as the symmetric representation $t=S$ of 
the permutation group $S_{3}$. The levels are grouped into oscillator shells
characterized by $n$. The levels belonging to an oscillator shell $n$ are 
further classified by $\sigma=n,n-2,\ldots,1$ or $0$ for $n$ odd or even. 
The quantum number $\sigma$ labels the representations of $SO(6)$, a subgroup 
of $U(6)$. The ground state has $n=0$ and $L^{P}=0^{+}$. We note, that the 
$n=\sigma =1$ shell is absent, since it does not contain a symmetric state 
with $t=S$. The two-phonon multiplet $n=2$ consists of the states 
$L^{P}=2^{+}$ with $\sigma =2$ and $0^{+}$ with $\sigma =0$. 
The degeneracy of the harmonic oscillator shells can be split 
by adding invariants of subgroups of $U(6)$ \cite{BL2}. 

\begin{figure}
\centering
\setlength{\unitlength}{1.0pt} 
\begin{picture}(260,220)(0,0)
\thinlines
\put (  0,  0) {\line(1,0){260}}
\put (  0,220) {\line(1,0){260}}
\put (  0,  0) {\line(0,1){220}}
\put (260,  0) {\line(0,1){220}}
\thicklines
\put ( 50, 30) {\line(1,0){20}}
\put ( 50, 90) {\line(1,0){20}}
\put ( 50,120) {\line(1,0){20}}
\put ( 50,150) {\line(1,0){20}}
\put (140, 90) {\line(1,0){20}}
\put (140,150) {\line(1,0){20}}
\put (190,150) {\line(1,0){20}}
\thinlines
\put ( 25,180) {$n \backslash \sigma$}
\put ( 60,180) {$n$}
\put ( 30, 27) {$0$}
\put ( 30, 87) {$2$}
\put ( 30,117) {$3$}
\put ( 30,147) {$4$}
\put ( 75, 27) {$0^+$}
\put ( 75, 87) {$2^+$}
\put ( 75,117) {$3^-,\,1^-$}
\put ( 75,147) {$4^+,\,2^+,\,0^+$}
\put (140,180) {$n-2$}
\put (165, 87) {$0^+$}
\put (165,147) {$2^+$}
\put (190,180) {$n-4$}
\put (215,147) {$0^+$}
\put (150, 25) {$U(6) \supset SO(6)$}
\end{picture}
\vspace{15pt}
\caption[]{Schematic spectrum of the harmonic oscillator with $U(6) \supset
SO(6)$ symmetry. The number of bosons is $N=4$. All states are symmetric
under $S_3$.}
\label{hosc}
\end{figure}

The classical limit of the $U(6)$ Hamiltonian is given by 
\ba
H_{1,{\rm cl}} &=& \frac{1}{N} \langle N;\vec{\alpha}_{\rho },\vec{\alpha}%
_{\lambda }\mid \,: H_1 :\,\mid N;\vec{\alpha}_{\rho },\vec{\alpha}_{\lambda
}\rangle  \nonumber \\
&=& \epsilon_{1} \, (\vec{\alpha}_{\rho }\cdot \vec{\alpha}%
_{\rho }^{\,\ast }+\vec{\alpha}_{\lambda }\cdot \vec{\alpha}_{\lambda
}^{\,\ast })  \nonumber \\
&=& \epsilon_{1} \frac{1}{2}\left( p_{\rho }^{2}+q_{\rho}^{2} 
+ L_{\rho }^{2}/q_{\rho}^{2}+p_{\lambda }^{2}+q_{\lambda }^{2} 
+ L_{\lambda }^{2}/q_{\lambda}^{2}\right) ~,
\ea
where $L_{k}^{2}$ is the angular momentum in polar coordinates 
\ba
L_{k}^{2} &=& p_{\theta_{k}}^{2} 
+ \frac{p_{\phi_{k}}^{2}}{\sin^{2}\theta_{k}} ~,
\ea
with $k=\rho$, $\lambda$. A change of variables to the hyperspherical 
coordinates $q$ and $\chi$ 
\ba
q_{\rho} \;=\; q \sin \chi ~, \hspace{1cm} q_{\lambda} \;=\; q \cos \chi ~,
\label{hyper}
\ea
and their conjugate momenta, $p$ and $p_{\chi}$, reduces the classical 
limit to 
\ba
H_{1,{\rm cl}} &=& \epsilon_{1} \frac{1}{2}\left( p^{2} + q^2 
+ \frac{\Lambda^2}{q^2} \right) ~. 
\label{hcl1}
\ea
Here $\Lambda^2$ is the angular momentum for rotations in six dimensions 
\ba
\Lambda^2 &=& p_{\chi}^{2} + \frac{L_{\rho}^2}{\sin^2 \chi}  
+ \frac{L_{\lambda}^2}{\cos^2 \chi} ~.
\label{angmom6}
\ea
It is the classical limit of the $SO(6)$ Casimir operator and is a constant 
of the motion. Therefore, we can first apply the requantization conditions 
to the coordinates and momenta contained in $\Lambda^2$, which yields that  
$\Lambda^2$ be replaced by $\sigma^2/N^2$ \cite{onno}. The difference from 
the exact result $\sigma(\sigma+4)/N^2$ is typical for the semi-classical 
approximation. The remaining quantization condition in the $(p,q)$ phase 
space is of the Bohr-Sommerfeld type
\ba
N \oint p dq \;=\; 2N \int dq \sqrt{\frac{2E}{N\epsilon_1}-q^2-
\frac{\sigma^2}{N^2q^2}} \;=\; 2\pi n_q ~.
\ea
The integral can be solved exactly to obtain 
\ba
E &=& \epsilon_1 \, (2n_q+\sigma) ~, 
\ea
which is identical to the exact result of Eq.~(\ref{e1} with $n=2n_q+\sigma$. 
This semi-classical analysis confirms the interpretation of the $U(6)$ 
limit of the ACM in terms of a six-dimensional spherical oscillator. 

\subsection{Dynamical symmetries: the SO(7) limit}

For the six-dimensional spherical oscillator, the number of oscillator
quanta $n$ is a good quantum number. However, when $v_{0}\neq 0$ in Eq.~(\ref
{hs3}), the oscillator shells with $\Delta n=\pm 2$ are mixed, and the
eigenfunctions are spread over many different oscillator shells. A dynamical 
symmetry that involves the mixing between oscillator shells, is provided by
the reduction 
\ba
U(7) \supset SO(7) \supset SO(6) \supset \cdots 
\ea
We consider now a dipole-dipole interaction which can be rewritten as the 
difference between the Casimir operators of $SO(7)$ and $SO(6)$
\ba
H_2 &=& \kappa \left[ \hat{N}(\hat{N}+5) - \hat{C}_{2SO(7)} 
+ \hat{C}_{2SO(6)} \right] 
\nonumber\\
&=& \kappa \left[ ( s^{\dagger} s^{\dagger}-b_{\rho}^{\dagger} \cdot
b_{\rho}^{\dagger}-b_{\lambda}^{\dagger} \cdot b_{\lambda}^{\dagger}) 
\, ( \tilde{s}\tilde{s}-\tilde{b}_{\rho}\cdot \tilde{b}_{\rho} - 
\tilde{b}_{\lambda }\cdot \tilde{b}_{\lambda }) + \hat{C}_{2SO(6)}\right] ~,  
\label{ham2}
\ea
where $\hat N$ is the number operator of Eq.~(\ref{number}). 
The energy spectrum in this case, called the $SO(7)$ limit, is given by 
the eigenvalues of the Casimir operators as 
\ba
E_2 &=& \kappa \, [(N-\omega)(N+\omega +5) 
+\sigma (\sigma+4) ] ~. 
\label{e2}
\ea
The label $\omega =N,N-2,\ldots,1$ or $0$ for $N$ odd or even, respectively, 
characterizes the symmetric representations of $SO(7)$, and 
$\sigma=0,2,3,\ldots,\omega$ those of $SO(6)$ (note that $\sigma=1$ 
is missing, since it does not contain a symmetric state). 

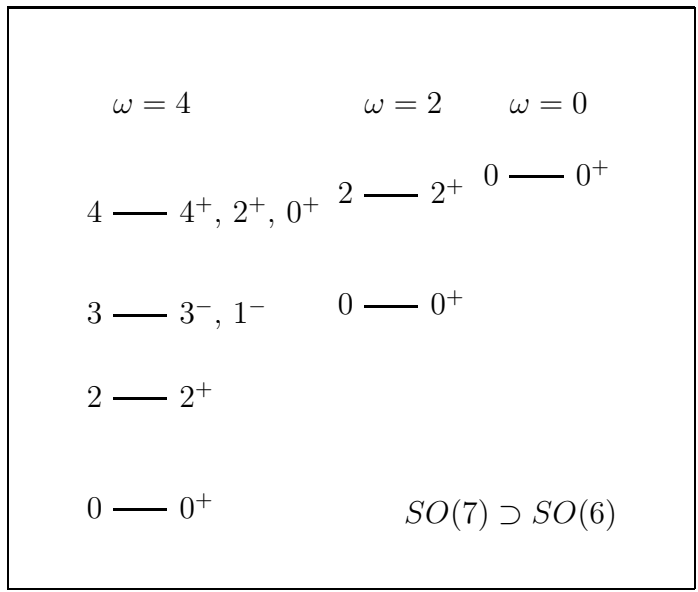
\begin{figure}
\centering
\setlength{\unitlength}{1.0pt} 
\begin{picture}(260,220)(0,0)
\thinlines
\put (  0,  0) {\line(1,0){260}}
\put (  0,220) {\line(1,0){260}}
\put (  0,  0) {\line(0,1){220}}
\put (260,  0) {\line(0,1){220}}
\thicklines
\put ( 40, 30) {\line(1,0){20}}
\put ( 40, 72) {\line(1,0){20}}
\put ( 40,103.5) {\line(1,0){20}}
\put ( 40,142) {\line(1,0){20}}
\put (135,107) {\line(1,0){20}}
\put (135,149) {\line(1,0){20}}
\put (190,156) {\line(1,0){20}}
\thinlines
\put ( 40,180) {$\omega=4$}
\put ( 30, 27) {$0$}
\put ( 30, 69) {$2$}
\put ( 30,100.5) {$3$}
\put ( 30,139) {$4$}
\put ( 65, 27) {$0^+$}
\put ( 65, 69) {$2^+$}
\put ( 65,100.5) {$3^-,\,1^-$}
\put ( 65,139) {$4^+,\,2^+,\,0^+$}
\put (135,180) {$\omega=2$}
\put (125,104) {$0$}
\put (125,146) {$2$}
\put (160,104) {$0^+$}
\put (160,146) {$2^+$}
\put (190,180) {$\omega=0$}
\put (180,153) {$0$}
\put (215,153) {$0^+$}
\put (150, 25) {$SO(7) \supset SO(6)$}
\end{picture}
\vspace{15pt}
\caption[]{Schematic spectrum of a deformed oscillator with $SO(7)$ symmetry.
The number of bosons is $N=4$. All states are symmetric under $S_3$.}
\label{defosc}
\end{figure}

In Fig.~\ref{defosc} we show the spectrum of the deformed oscillator with 
$SO(7)$ symmetry. The states are now ordered in bands labeled by $\omega$,
rather than in harmonic oscillator shells. Although the size of the model
space, and hence the total number of states, is the same as for the harmonic
oscillator, the ordering and classification of the states is different. For
example, in the $U(6)$ limit all states are vibrational, whereas the $SO(7)$
limit gives rise to a rotational-vibrational spectrum, characterized by a
series of rotational bands which are labeled by $\omega $, or equivalently
by the vibrational quantum number $v=(N-\omega )/2=0,1,\ldots $~.

The classical limit of the $SO(7)$ limit is given by 
\ba
H_{2,{\rm cl}} &=& \kappa (N-1) \left[ q^{2}p^{2} + (1-q^{2})^{2} 
+ \Lambda^2 \right] ~.
\ea
Here we have used the hyperspherical variables that were introduced in 
Eq.~(\ref{hyper}). Also in this case, the six-dimensional angular momentum 
$\Lambda^2$ is a constant of the motion, and hence can be requantized first. 
The remaining quantization condition in the $(p,q)$ phase space 
\ba
N \oint p dq \;=\; 2N \int dq \frac{1}{q} 
\sqrt{\frac{E-\kappa(N-1)\sigma^2/N}{\kappa N(N-1)}-(1-q^2)^2} 
\;=\; 2\pi v ~,
\ea
can be solved exactly to obtain 
\ba
E \;=\; 4 \kappa N \, v \left( 1-\frac{v}{N} \right) 
\left( 1-\frac{1}{N} \right) 
+ \kappa \, \sigma^2 \left( 1-\frac{1}{N} \right) ~.
\ea
In the large $N$ limit, this expression reduces to the exact one of 
Eq.~(\ref{e2}), if we associate the vibrational quantum number $v$ with 
$(N-\omega )/2$ 
\ba
E_2 \;=\; 4\kappa N \, v \left( 1-\frac{2v-5}{2N} \right) 
+ \kappa \, \sigma(\sigma +4) ~.
\label{ener}
\ea
To leading order in $N$, the frequency of the vibrational motion coincides. 
This analysis shows the connection between the $SO(7)$ dynamical symmetry 
and the deformed oscillator.

\subsection{Oblate symmetric top}

\begin{figure}
\setlength{\unitlength}{0.8pt}
\begin{picture}(450,200)(-50,0)
\thinlines
\put ( 60,  0) {$v_1$}
\put ( 25, 50) {\circle*{5}}
\put (125, 50) {\circle*{5}}
\put ( 75,150) {\circle*{5}}
\put ( 25, 50) {\line ( 1,0){100}}
\put ( 25, 50) {\line ( 1,2){ 50}}
\put (125, 50) {\line (-1,2){ 50}}
\multiput (  5, 35)( 5, 0){29}{\circle*{0.1}}
\multiput ( 75,150)( 0,-5){23}{\circle*{0.1}}
\thicklines
\put ( 75,150) {\vector( 0, 1){25}}
\put ( 25, 50) {\vector(-4,-3){20}}
\put (125, 50) {\vector( 4,-3){20}}
\thinlines
\put (210,  0) {$v_{2a}$}
\put (175, 50) {\circle*{5}}
\put (275, 50) {\circle*{5}}
\put (225,150) {\circle*{5}}
\put (175, 50) {\line ( 1,0){100}}
\put (175, 50) {\line ( 1,2){ 50}}
\put (275, 50) {\line (-1,2){ 50}}
\multiput (195, 35)( 5, 0){13}{\circle*{0.1}}
\multiput (225,150)( 0,-5){23}{\circle*{0.1}}
\thicklines
\put (225,150) {\vector( 0, 1){25}}
\put (175, 50) {\vector( 4,-3){20}}
\put (275, 50) {\vector(-4,-3){20}}
\thinlines
\put (360,  0) {$v_{2b}$}
\put (325, 50) {\circle*{5}}
\put (425, 50) {\circle*{5}}
\put (375,150) {\circle*{5}}
\put (325, 50) {\line ( 1,0){100}}
\put (325, 50) {\line ( 1,2){ 50}}
\put (425, 50) {\line (-1,2){ 50}}
\multiput (315, 30)(   5,2){21}{\circle*{0.1}}
\multiput (365, 50)(1.75,5){21}{\circle*{0.1}}
\thicklines
\put (375,150) {\vector( 1, 0){25}}
\put (325, 50) {\vector(-1,-2){10}}
\put (425, 50) {\vector(-1, 2){10}}
\end{picture}
\vspace{15pt}
\caption[]{Vibrations of an oblate top.}
\label{obltop}
\end{figure}
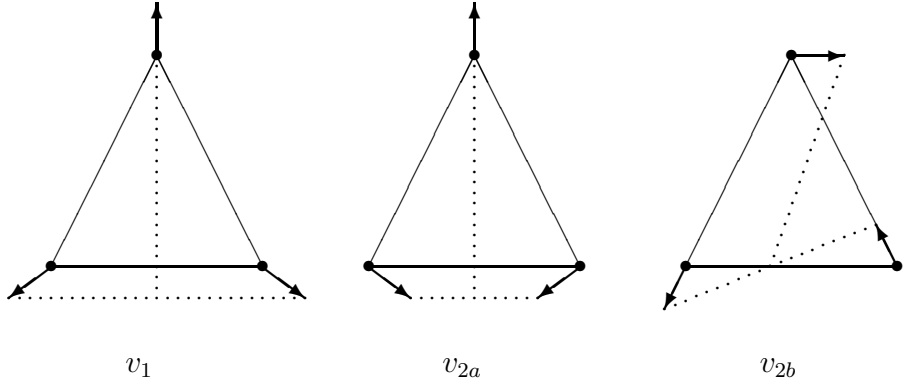

The potential energy surfaces of the $U(6)$ and $SO(7)$ limits only depend 
on the hyperspherical radius $q$. The corresponding equilibrium shapes are 
characterized by $q_0=0$ and $q_0=1$, respectively. Another interesting case 
is provided by the Hamiltonian \cite{ACM} 
\ba
H_{3,{\rm vib}} &=&\xi _{1}\,(R^{2}\,s^{\dagger }s^{\dagger }-b_{\rho
}^{\dagger }\cdot b_{\rho }^{\dagger }-b_{\lambda }^{\dagger }\cdot
b_{\lambda }^{\dagger })\,(R^{2}\,\tilde{s}\tilde{s}-\tilde{b}_{\rho }\cdot 
\tilde{b}_{\rho }-\tilde{b}_{\lambda }\cdot \tilde{b}_{\lambda })  
\nonumber\\
&&+\xi _{2}\,\left[ (b_{\rho }^{\dagger }\cdot b_{\rho }^{\dagger
}-b_{\lambda }^{\dagger }\cdot b_{\lambda }^{\dagger })\,(\tilde{b}_{\rho
}\cdot \tilde{b}_{\rho }-\tilde{b}_{\lambda }\cdot \tilde{b}_{\lambda
})+4\,(b_{\rho }^{\dagger }\cdot b_{\lambda }^{\dagger })\,(\tilde{b}%
_{\lambda }\cdot \tilde{b}_{\rho })\right] ~.  \label{oblate}
\ea
For $R^{2}=0$, this Hamiltonian has $U(7)\supset U(6)$ symmetry and
corresponds to a spherical vibrator, whereas for $R^{2}=1$ and $\xi_{2}=0$
it has $U(7)\supset SO(7)$ symmetry and corresponds to a deformed 
oscillator. The general case with $R^{2}\neq 0$ and $\xi_{1}$, $\xi_{2}>0$ 
does not correspond to a dynamical symmetry, and hence its energy spectrum 
cannot be obtained in closed analytic form. In this case, the energy 
eigenvalues and eigenvectors are calculated numerically by diagonalizing 
the Hamiltonian in an appropriate basis. 
However, an approximate energy formula can still be 
derived in a semiclassical mean-field analysis. The general expression of 
the classical limit of the Hamiltonian of Eq.~(\ref{oblate}) has a 
complicated structure. We first study the potential energy surface which is 
obtained by setting all momenta equal to zero. Its equilibrium configuration 
is given by 
\ba
q_{0} \;=\; \sqrt{2R^{2}/(1+R^{2})} ~, \hspace{1cm} 
\chi_{0} \;=\; \pi/4 ~, \hspace{1cm} \zeta_{0} \;=\; \pi/4 ~,
\ea
where $2\zeta$ denotes the relative angle between $\vec{\alpha}_{\rho}$ 
and $\vec{\alpha}_{\lambda}$. In the limit of small oscillations around the 
minimum $q=q_{0}+\Delta q$, $\chi =\chi_{0}+\Delta \chi$ and 
$\zeta =\zeta_{0}+\Delta \zeta$, the intrinsic degrees of freedom $q$, $\chi$ 
and $\zeta$ decouple and become harmonic. As a result we find that the 
classical limit, to leading order in $N$, is given by 
\ba
H_{3,{\rm cl}} &=& \xi_{1}N \left[ \frac{2R^{2}}{1+R^{2}}p^{2} 
+ 2R^{2}(1+R^{2})(\Delta q)^{2} \right] 
\nonumber\\
&& +\xi_{2}N \left[ 
 p_{\chi}^{2}+\frac{4R^{4}}{(1+R^{2})^{2}}(\Delta \chi)^{2} 
+p_{\zeta}^{2}+\frac{4R^{4}}{(1+R^{2})^{2}}(\Delta \zeta)^{2}\right] ~.
\ea
Standard requantization of the harmonic oscillator yields the vibrational
energy spectrum of an oblate top 
\ba
E_{3,{\rm vib}} \;=\; \omega_{1}(v_{1}+\frac{1}{2}) 
+ \omega_{2}(v_{2}+1) ~, 
\label{e3vib}
\ea
with frequencies 
\ba
\omega _{1} \;=\; 4NR^{2} \xi_{1} ~,  \hspace{1cm} 
\omega _{2} \;=\; \frac{4NR^{2}}{1+R^{2}} \xi_{2} ~,
\ea
which in agreement with the results obtained in a normal mode analysis 
\cite{BIL}. Here $v_{1}$ represents the vibrational quantum number for 
a symmetric stretching vibration, and $v_2$ for a degenerate doublet of 
an antisymmetric stretching $(v_{2a})$ and a bending $(v_{2b})$ vibration 
(see Fig.~\ref{obltop}). The vibrational excitations can be labeled by 
$(v_1,v_2^{\ell})$ with $\ell=v_{2},v_{2}-2,\ldots,1$ or $0$ for 
$v_{2}=v_{2a}+v_{2b}$ odd or even, respectively.  

Next we consider the rotational Hamiltonian 
\ba
H_{3,{\rm rot}} &=& 2\kappa_1 \, (b_{\rho }^{\dagger} \times \tilde{b}_{\rho} 
+ b_{\lambda}^{\dagger} \times \tilde{b}_{\lambda})^{(1)} \cdot 
(b_{\rho}^{\dagger} \times \tilde{b}_{\rho}+b_{\lambda}^{\dagger} \times 
\tilde{b}_{\lambda})^{(1)}  \nonumber\\
&&+3\kappa_2 \, (b_{\rho}^{\dagger} \times \tilde{b}_{\lambda}
-b_{\lambda}^{\dagger} \times \tilde{b}_{\rho}) ^{(0)} \cdot 
(b_{\lambda}^{\dagger} \times \tilde{b}_{\rho}-b_{\rho}^{\dagger} \times 
\tilde{b}_{\lambda})^{(0)} ~. 
\ea
Both terms commute with the general $S_{3}$ invariant Hamiltonian of 
Eq.~(\ref{hs3}), and hence correspond to exact symmetries \cite{BIL}. 
The eigenvalues are given by 
\ba
E_{3,{\rm rot}} &=& \kappa_{1} \, L(L+1) + \kappa_{2} \, m_{F}^{2} 
\nonumber\\
&=& \kappa_{1} \, L(L+1) + \kappa_{2} \, (K \mp 2\ell)^{2} ~.
\label{e3rot}
\ea
Here we have used, that for the oblate top the quantum number $m_{F}$ is 
related to the projection $K$ of the angular momentum on the symmetry-axis 
and the value of $\ell$ \cite{BDL}. 
The last term contains the effects of the Coriolis force which gives rise to
a $8\kappa_{2} K \ell$ splitting of the rotational levels.

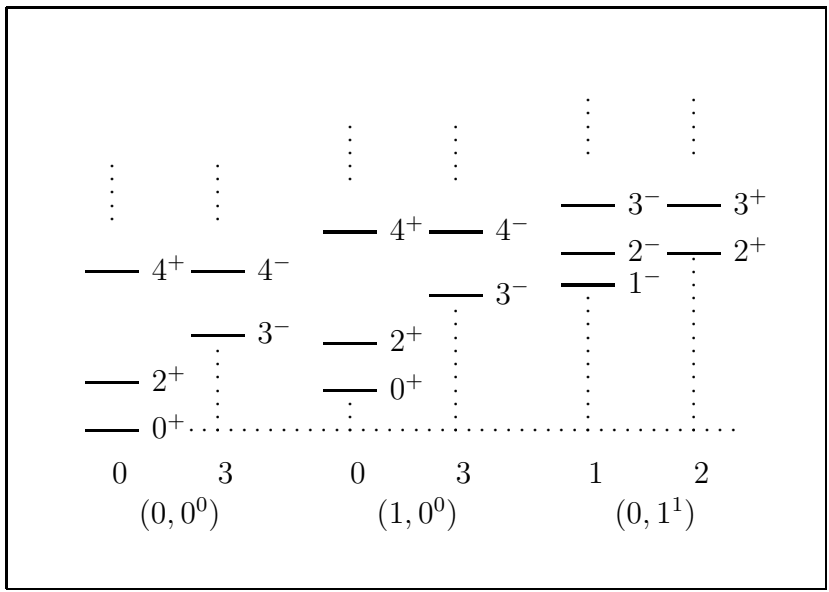
\begin{figure}
\centering
\setlength{\unitlength}{1.0pt} 
\begin{picture}(310,220)(0,0)
\thinlines
\put (  0,  0) {\line(1,0){310}}
\put (  0,220) {\line(1,0){310}}
\put (  0,  0) {\line(0,1){220}}
\put (310,  0) {\line(0,1){220}}
\thicklines
\put ( 30, 60) {\line(1,0){20}}
\put ( 30, 78) {\line(1,0){20}}
\put ( 30,120) {\line(1,0){20}}
\put ( 70, 96) {\line(1,0){20}}
\put ( 70,120) {\line(1,0){20}}
\multiput ( 40,140)(0,5){5}{\circle*{0.1}}
\multiput ( 80,140)(0,5){5}{\circle*{0.1}}
\multiput ( 70, 60)(5,0){42}{\circle*{0.1}}
\multiput ( 80, 60)(0,5){7}{\circle*{0.1}}
\thinlines
\put ( 40, 40) {$0$}
\put ( 80, 40) {$3$}
\put ( 50, 25) {$(0,0^0)$}
\put ( 55, 57) {$0^+$}
\put ( 55, 75) {$2^+$}
\put ( 55,117) {$4^+$}
\put ( 95, 93) {$3^-$}
\put ( 95,117) {$4^-$}
\thicklines
\put (120, 75) {\line(1,0){20}}
\put (120, 93) {\line(1,0){20}}
\put (120,135) {\line(1,0){20}}
\put (160,111) {\line(1,0){20}}
\put (160,135) {\line(1,0){20}}
\multiput (130,155)(0,5){5}{\circle*{0.1}}
\multiput (170,155)(0,5){5}{\circle*{0.1}}
\multiput (130, 60)(0,5){3}{\circle*{0.1}}
\multiput (170, 60)(0,5){10}{\circle*{0.1}}
\thinlines
\put (130, 40) {$0$}
\put (170, 40) {$3$}
\put (140, 25) {$(1,0^0)$}
\put (145, 72) {$0^+$}
\put (145, 90) {$2^+$}
\put (145,132) {$4^+$}
\put (185,108) {$3^-$}
\put (185,132) {$4^-$}
\thicklines
\put (210,115) {\line(1,0){20}}
\put (210,127) {\line(1,0){20}}
\put (210,145) {\line(1,0){20}}
\put (250,127) {\line(1,0){20}}
\put (250,145) {\line(1,0){20}}
\multiput (220,165)(0,5){5}{\circle*{0.1}}
\multiput (260,165)(0,5){5}{\circle*{0.1}}
\multiput (220, 60)(0,5){11}{\circle*{0.1}}
\multiput (260, 60)(0,5){14}{\circle*{0.1}}
\thinlines
\put (220, 40) {$1$}
\put (260, 40) {$2$}
\put (230, 25) {$(0,1^1)$}
\put (235,112) {$1^-$}
\put (235,124) {$2^-$}
\put (235,142) {$3^-$}
\put (275,124) {$2^+$}
\put (275,142) {$3^+$}
\end{picture}
\vspace{1cm}
\caption[]{Schematic spectrum of an oblate symmetric top. The rotational 
bands are labeled by $K$ and $(v_1,v_2^{\ell})$ (bottom). 
All states are symmetric $t=S$ under $S_3$.} 
\label{top}
\end{figure}

In Fig.~\ref{top} we show the structure of the spectrum of the oblate top 
according to the approximate energy formula of Eqs.~(\ref{e3vib}) and 
(\ref{e3rot}). The energy spectrum consists of a series of rotational bands 
labeled by $(v_1,v_2^{\ell})$ and $K$. The degeneracy between states with 
different values of $K$ can be split by the last term in Eq.~(\ref{e3rot}). 
The vibrational bands with $(v_1,0^{0})$ can have angular momenta and parity 
$L^{P}=0^{+},2^{+},3^{-},4^{\pm },\ldots$~, whereas the angular momentum 
content of the doubly degenerate vibrations $(v_1,1^{1})$ is given by 
$L^{P}=1^{-},2^{\mp },3^{\mp},\ldots$~. 

\section{The nucleus $^{12}$C} 

As an application, we investigate the extent to which properties of the 
low-lying spectrum of $^{12}$C can be described in terms of the ACM. 
The ACM provides a set of explicit formulas for energies, electromagnetic 
transition rates and form factors that can be easily be compared with 
experiments for any of the three special solutions discussed in the previous 
sections. The differences between the $U(6)$ limit, the $SO(7)$ limit and 
the oblate top, are most pronounced for the form factors \cite{ACM}. The 
experimental data for  
the elastic form factor $|{\cal F}(0^+_1 \rightarrow 0^+_1;q)|^2$ and the 
transition form factor $|{\cal F}(0^+_1 \rightarrow 0^+_2;q)|^2$ 
of $^{12}$C show a clear minimum. Only the oblate top limit can account for 
these qualitative features. In the $U(6)$ limit the elastic form 
factor falls off exponentially and has no minimum, whereas in the $SO(7)$ 
limit the inelastic form factor vanishes identically. Therefore, we analyze 
the spectroscopy of $^{12}$C in the oblate top limit of the ACM. 

\begin{figure}
\centering
\vspace{15pt}
\setlength{\unitlength}{0.8pt}
\begin{picture}(520,250)(-60,30)
\thinlines
\put (  0, 35) {\line(0,1){215}}
\put (230, 35) {\line(0,1){215}}
\put (460, 35) {\line(0,1){215}}
\put (  0, 35) {\line(1,0){460}}
\put (  0,250) {\line(1,0){460}}
\thicklines
\put (  0, 60) {\line(1,0){5}}
\put (  0,110) {\line(1,0){5}}
\put (  0,160) {\line(1,0){5}}
\put (  0,210) {\line(1,0){5}}
\put (225, 60) {\line(1,0){10}}
\put (225,110) {\line(1,0){10}}
\put (225,160) {\line(1,0){10}}
\put (225,210) {\line(1,0){10}}
\put (455, 60) {\line(1,0){5}}
\put (455,110) {\line(1,0){5}}
\put (455,160) {\line(1,0){5}}
\put (455,210) {\line(1,0){5}}
\put (-20, 55) { 0}
\put (-20,105) { 5}
\put (-20,155) {10}
\put (-20,205) {15}
\put (-60,180) {E(MeV)}
\put (170, 55) {EXP}
\put (400, 55) {TH}
\put ( 20, 60.0) {\line(1,0){20}}
\put ( 20,104.4) {\line(1,0){20}}
\put ( 20,156.4) {\line(1,0){20}}
\put ( 20,200.8) {\line(1,0){20}}
\thinlines
\put ( 45, 55.0) {$0^+$}
\put ( 45, 99.4) {$2^+$}
\put ( 45,151.4) {$3^-$}
\put ( 45,195.8) {$4^+$}
\thicklines
\put ( 70,136.5) {\line(1,0){20}}
\put ( 70,171.6) {\line(1,0){20}}
\thinlines
\put ( 95,131.5) {$0^+$}
\put ( 95,168.6) {$(2^+)$}
\thicklines
\put (120,163.0) {\line(1,0){20}}
\thinlines
\put (145,158.0) {$(0^+)$}
\thicklines
\put (170,168.4) {\line(1,0){20}}
\put (170,178.3) {\line(1,0){20}}
\thinlines
\put (195,163.4) {$1^-$}
\put (195,178.3) {$2^-$}
\thicklines
\put (250, 60.0) {\line(1,0){20}}
\put (250,102.4) {\line(1,0){20}}
\put (250,156.3) {\line(1,0){20}}
\put (250,201.4) {\line(1,0){20}}
\thinlines
\put (275, 55.0) {$0^+$}
\put (275, 97.4) {$2^+$}
\put (275,151.3) {$3^-$}
\put (275,196.4) {$4^+$}
\thicklines
\put (300,136.5) {\line(1,0){20}}
\put (300,149.4) {\line(1,0){20}}
\thinlines
\put (325,131.5) {$0^+$}
\put (325,144.4) {$2^+$}
\thicklines
\put (350,182.2) {\line(1,0){20}}
\thinlines
\put (375,177.2) {$0^+$}
\thicklines
\put (400,168.4) {\line(1,0){20}}
\put (400,204.3) {\line(1,0){20}}
\put (400,195.7) {\line(1,0){20}}
\put (400,207.7) {\line(1,0){20}}
\thinlines
\put (425,163.4) {$1^-$}
\put (425,197.3) {$2^-$}
\put (425,185.7) {$2^+$}
\put (425,209.7) {$3^+$}
\end{picture}
\vspace{15pt}
\caption[]{Comparison between the low-lying experimental spectrum of $^{12}$C 
and that calculated with the oblate top Hamiltonian with $N=10$.}
\label{ec12}
\end{figure}
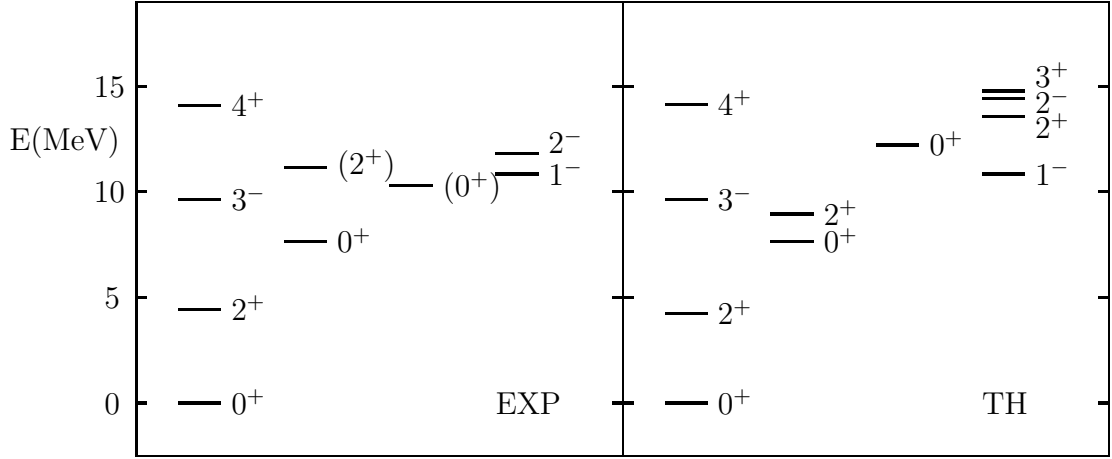

The oblate top Hamiltonian is given by Eqs.~(\ref{e3vib}) and (\ref{e3rot}). 
The coefficients $\xi_1$, $\xi_2$, $\kappa_1$ and $\kappa_2$ are determined 
in a fit to the excitation energies of $^{12}$C \cite{ACM}. 
The number of bosons is 
taken to be $N=10$. In Fig.~\ref{ec12} we show a comparison between the 
experimental data and the calculated states of the oblate top with energies 
$< 15$ MeV. One can clearly identify in the experimental spectrum the states 
$0^{+}$, $2^{+}$, $3^{-}$, $4^{+}$ of the ground rotational band, the first 
$0^{+}$ state of the stretching vibration and the first $1^{-}$ state of the 
bending vibration. 

\begin{figure}

\vfill 

\begin{minipage}{.5\linewidth}
\centerline{\epsfig{file=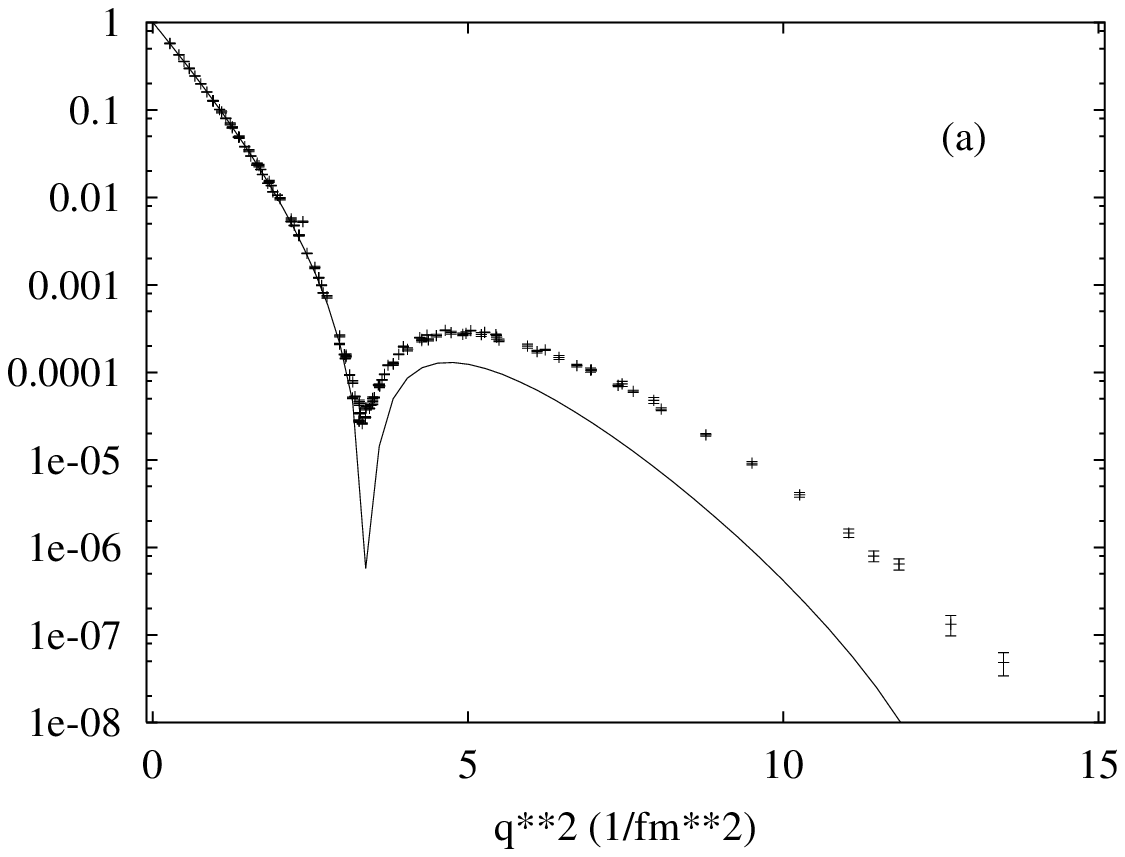,width=\linewidth}}
\end{minipage}\hfill
\begin{minipage}{.5\linewidth}
\centerline{\epsfig{file=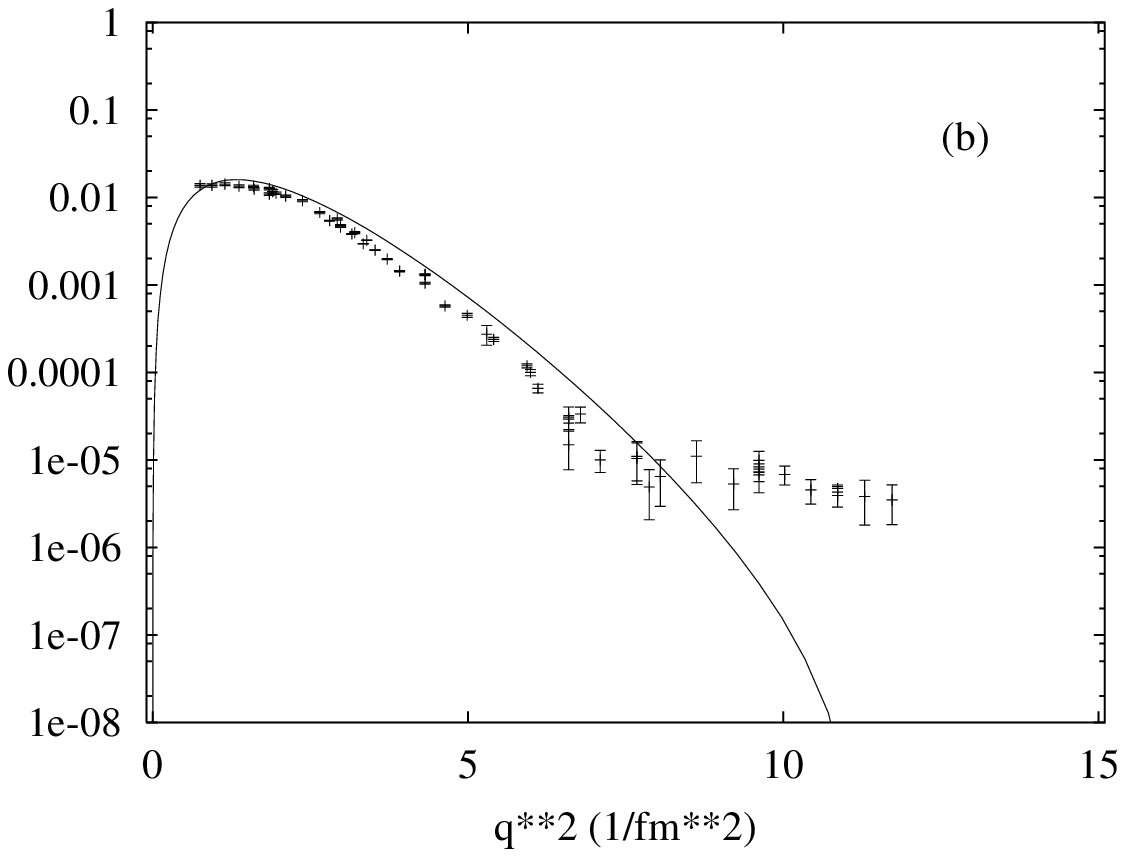,width=\linewidth}}
\end{minipage}

\begin{minipage}{.5\linewidth}
\centerline{\epsfig{file=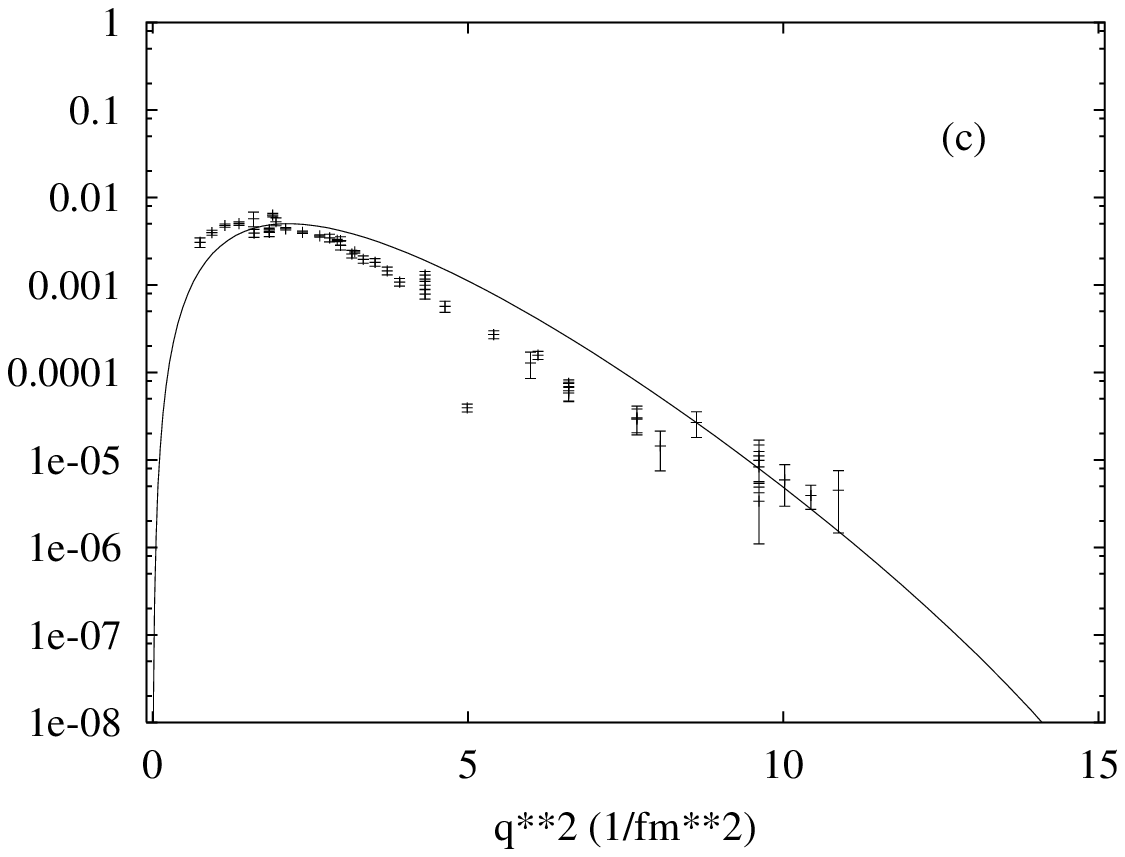,width=\linewidth}}
\end{minipage}\hfill
\begin{minipage}{.5\linewidth}
\centerline{\epsfig{file=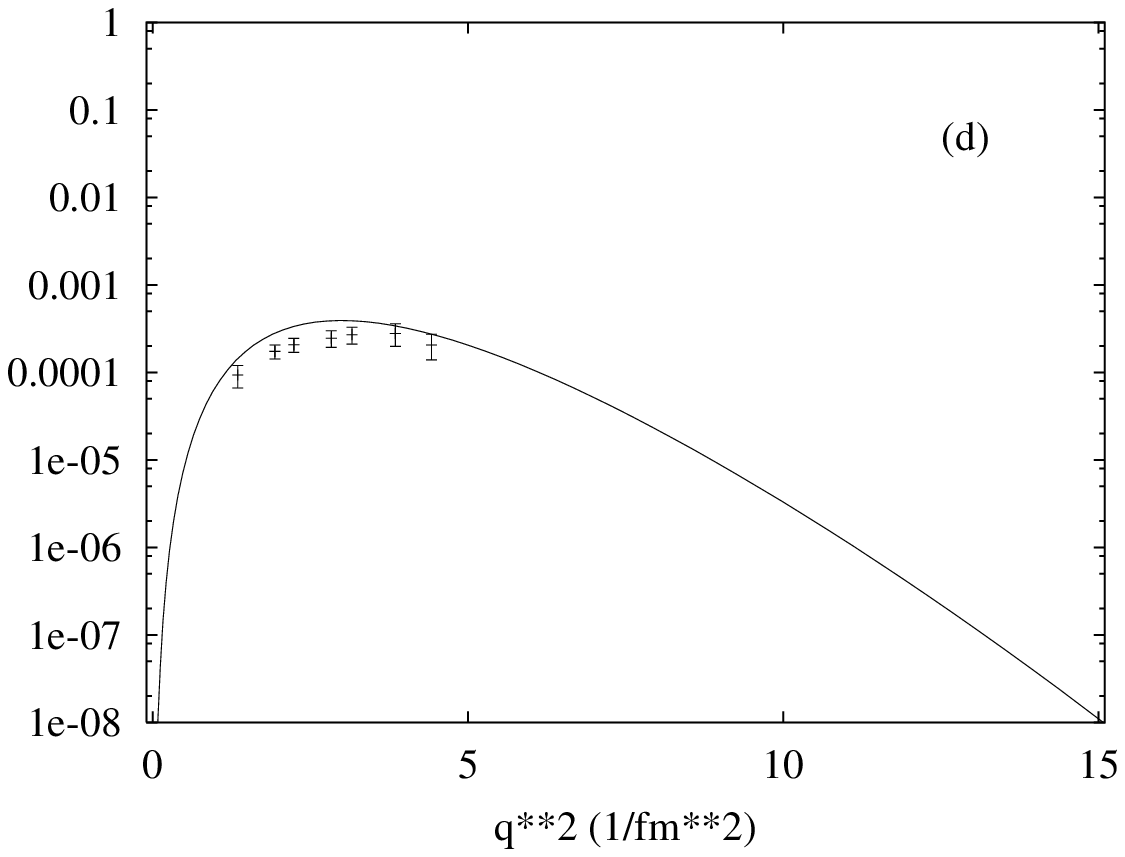,width=\linewidth}}
\end{minipage}

\begin{minipage}{.5\linewidth}
\centerline{\epsfig{file=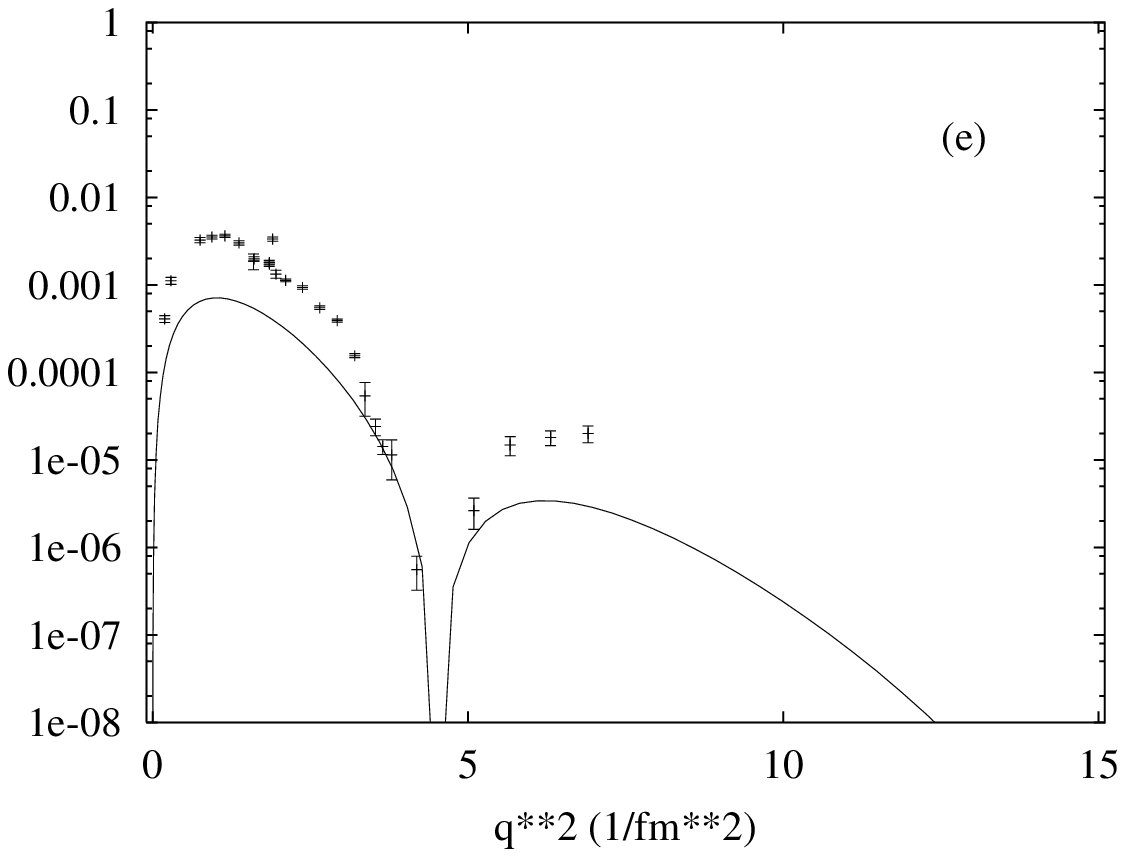,width=\linewidth}}
\end{minipage}\hfill
\begin{minipage}{.5\linewidth}
\centerline{\epsfig{file=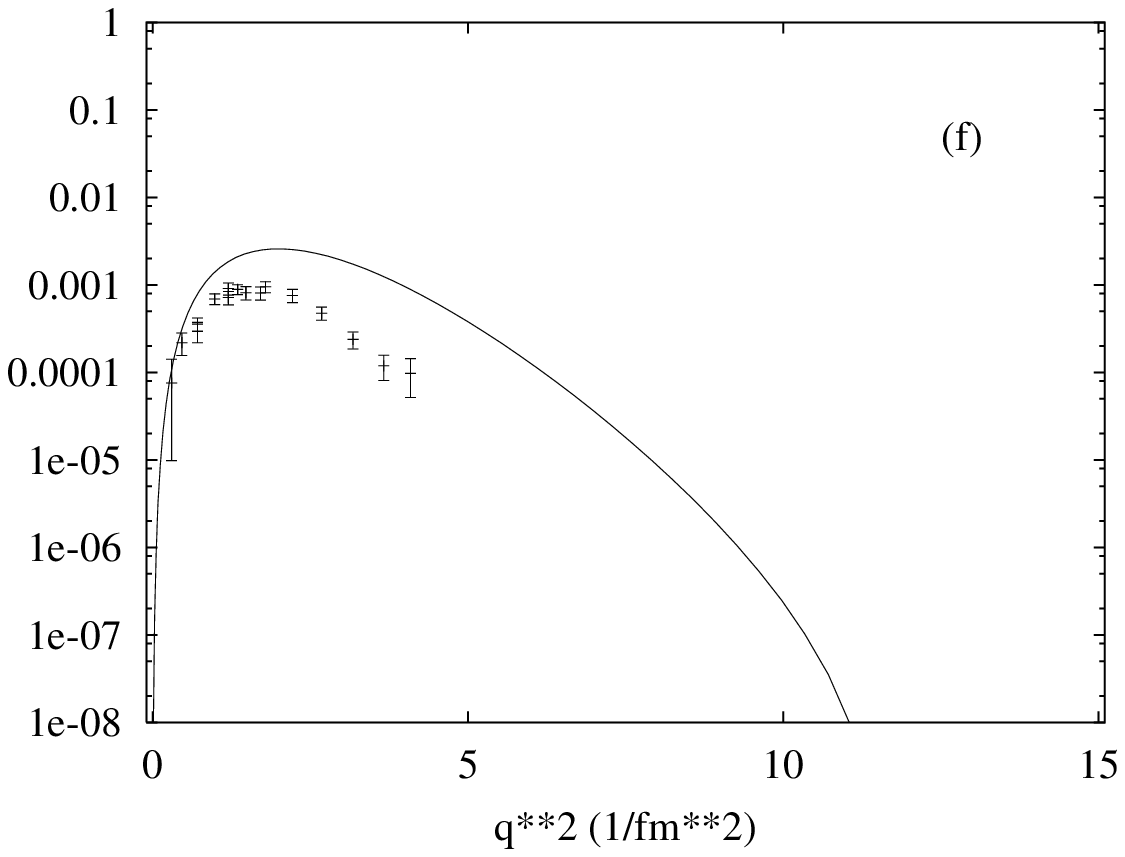,width=\linewidth}}
\end{minipage}

\caption[]{Comparison between the experimental form factors 
$|{\cal F}(0^+_1 \rightarrow L^P_i;q)|^2$ of $^{12}$C 
for the final states (a) $L^P_i=0^+_1$ (elastic), (b) $L^P_i=2^+_1$, 
(c) $L^P_i=3^-_1$, (d) $L^P_i=4^+_1$, (e) $L^P_i=0^+_2$, and 
(f) $L^P_i=1^-_1$ and those obtained for the oblate top with $N=10$.}
\label{ffc12}
\end{figure}

Form factors for electron scattering on $^{12}$C have been measured in 
\cite{reuter}-\cite{nakada}. In Fig.~\ref{ffc12} we show a 
comparison between experimental and theoretical form factors calculated 
in the oblate top limit of the ACM. The two coefficients that enter in the 
calculation of the theoretical form factors are determined by the minimum 
in the elastic form factor and the charge radius of $^{12}$C \cite{ACM}. 
The analysis of the experimental form factors appears to indicate that
the triangular configuration describes the data reasonably well for the 
rotational band $0_1^{+}$, $2_1^{+}$, $3_1^{-}$, $4_1^{+}$, although with 
large rotation-vibration interactions. The situation is different for the 
vibrational excitations $0_{2}^{+}$ and $1_{1}^{-}$. Here the shape of the 
form factors is well reproduced but its magnitude is not. 

\section{Summary and conclusions}

In this contribution, we discussed cluster configurations consisting of 
three identical particles. These configurations are relevant for both 
hadronic physics (baryons as clusters of three constituent quarks), molecular 
physics (X$_3$ molecules) and nuclear physics ($^{12}$C as a cluster 
of three $\alpha$ particles). It was suggested to treat the relative motion 
of the clusters in terms of the algebraic cluster model. The ACM is based on 
the algebraic quantization of the relative Jacobi variables which gives rise 
to a $U(7)$ spectrum generating algebra. We studied three special cases, 
for which the ACM provides a set of explicit formulas for energies, 
electromagnetic transition rates and form factors that can be easily be 
compared with experiments. In a semiclassical analysis it was shown that 
these three limits correspond to the spherical oscillator, the deformed 
oscillator and the oblate symmetric top. 

The latter case, the oblate top, was applied to the low-lying states 
of $^{12}$C. In particular, we investigated the transition form factors. 
The shape of the form factors is reproduced reasonably well, lending support 
to the interpretation of the states of $^{12}$C as rotational and 
vibrational excitations of a triangular configuration of three $\alpha$ 
particles. However, the discrepancies with the observed strengths implies 
a large mixing with other configurations, and possibly the need to include 
higher order rotation-vibration couplings in the Hamiltonian. 

Finally, we note that the ACM provides a general framework to study 
three-body clusters, which is not restricted to the case of three identical 
particles at the vertices of a triangle discussed in this contribution. It 
can be applied to other situations as well, such as nonidentical particles 
\cite{BL} and/or other geometrical configurations which may be relevant for 
a description of giant trinuclear molecules in ternary cold fission 
\cite{greiner,hess}. 

\section*{Acknowledgements}

This work was supported in part by CONACyT under project 
32416-E, and by DPAGA under project IN106400.

\end{document}